\documentstyle[11pt]{article}
\begin{document}



\def\e{{\hat e}}
\def\ie{{$i.e.$}}
\def\ra{{\rightarrow}}
\def\a{{\alpha}}
\def\b{{\beta}}
\def\eps{{\epsilon}}
\def\n{{\eta}}
\def\g{\gamma}
\def\s{{\sigma}}
\def\r{{\rho}}
\def\z{{\zeta}}
\def\x{{\xi}}
\def\d{{\delta}}
\def\t{{\theta}}
\def\l{{\lambda}}
\def\ca{{\cal A}}
\def\cd{{\cal D}}
\def\ce{{\cal E}}
\def\cg{{\cal G}}
\def\co{{\cal O}}
\def\cn{{\cal N}}
\def\cs{{\cal S}}
\def\cz{{\cal Z}}
\def\cv{{\nu}}
\def\ck{{\cal K}}
\def\pr{{\partial}}
\def\prt{{\pr}_{\t}}
\def\tri{{\triangle}}
\def\na{{\nabla }}
\def\S{{\Sigma}}
\def\G{{\Gamma }}
\def\sp{\vspace{.1in}}
\def\hs{\hspace{.25in}}

\newcommand{\be}{\begin{equation}} \newcommand{\ee}{\end{equation}}
\newcommand{\bea}{\begin{eqnarray}}\newcommand{\eea}
{\end{eqnarray}}


\begin{titlepage}

\begin{flushright}
{{Goteborg-ITP-99-10}\\{hep-th/9907117}\\{(Revised version)}}
\end{flushright}

\begin{center}
\baselineskip= 24 truept
\vspace{.5in}

{\Large \bf Generalized Dirichlet Branes and Zero-modes}

\vspace{.4in}
{\large Supriya Kar\footnote{supriya@fy.chalmers.se}}

\end{center}
\begin{center}
\baselineskip= 18 truept

{\large \it Institute of Theoretical Physics \\
Goteborg University and Chalmers University of Technology \\
S-412 96 Goteborg, Sweden}

\end{center}

\vspace{.4in}

\baselineskip= 14 truept

\begin{center}
{\bf Abstract}
\end{center}
\vspace{.25in}

We investigate the effective dynamics of an arbitrary Dirichlet p-brane,
in a path-integral formalism, by incorporating the massless
excitations of closed string modes in open bosonic 
string theory. It is shown that the closed string background fields
in the bosonic sector of type II theories induce
invariant extrinsic curvature on the world-volume. 
In addition, the curvature can be seen to be associated with a divergence
at the boundary of string world-sheet. 
The re-normalization of the collective coordinates,
next to leading order in its derivative expansion, is performed to
handle the divergence and the effective dynamics is encoded in 
Dirac-Born-Infeld action. 
Furthermore, the collective dynamics is generalized to
include appropriate fermionic partners in type I super-string theory.
The role of string modes is reviewed in terms of the collective coordinates
and the gauge theory on the world-volume is argued to be non-local in
presence of the $U(1)$ invariant field strength.

\vspace{.25in}

\thispagestyle{empty}
\end{titlepage}

\baselineskip= 18 truept

\section{Introduction}

In super-string theory, most of the non-perturbative insights 
have been analyzed by investigating various features of  
Dirichlet p-brane ($D_p$-brane), where `$p$' denotes the number of spatial
dimensions. Since strings are coupled to the
non-trivial background fields through conformal symmetry,
a systematic study \cite{notes} of the $D_p$-brane involves
arbitrary type II string backgrounds and have been in focus since its
inception as the Ramond-Ramond (RR) charge carriers \cite{polchinski}.
Among various aspects of the $D_p$-brane, one of the important is the
$D_p$-brane dynamics. Several investigations in this direction have
been discussed using different techniques
\cite{callan}. In general, the world-volume gauge
theory can be seen as an approximation to the 
underlying open string theory 
\cite{dailp,leigh,mrdouglas,tseytlin3,tseytlin4}.  In this context,
a path-integral formalism to deal with the collective dynamics of a
D-particle ($D_0$) \cite{kazama} and subsequently for a D-string ($D_1$)
\cite{karkazama,kar} were developed.

\sp
From a microscopic point of view, the $D_p$-brane is characterized by 
the $(p+1)$-dimensional hyper-surface with open string ending on it 
\cite{dailp}. The $U(1)$ gauge field associated with the ends of the
open string contribute towards the world-volume dynamics
of the $D_p$-brane. Apparently, the world-volume gauge field gives rise to 
the extra degrees of freedom unlike the Green-Schwarz (fundamental) string.
The extra gauge degrees of freedom are argued by Callan and Klebanov
\cite{callan} as the
massive states of the open string{\footnote{In this article, we show that the
gauge degrees of freedom is essentially responsible for 
the non-local description on the world-volume of the $D_p$-brane.}}.

\sp
To a tree level approximation in string theory, the
gauge theory on the world-volume of a $D_p$-brane is a Yang-Mills
theory. The next order in $\a'$ expansion gives rise to the 
Born-Infeld action \cite{leigh} and
describes the world-volume dynamics of the $D_p$-brane.
In general, the Born-Infeld action incorporates corrections 
to all orders in $\a'$ and the exact dynamics of a $D_p$-brane
still remains unanswered. In most of the calculations, bosonic
$D_p$-branes are considered for simplicity. In 
presence of background fields, $D_p$-brane can be seen to form a
bound state with other lower branes in the theory. In that case,
the $D_p$-brane becomes non BPS though the bulk theory (type II)
still preserves super-symmetry. In this context, the
super-symmetric formulation for the $D_p$-brane dynamics 
has been worked out \cite{cederwall} in great detail in type II theories.

\sp
In this paper, we extend our analysis \cite{kar}
for a D-string ($p=1$) 
in presence of closed string backgrounds to an arbitrary $D_p$-brane.
Now, the antisymmetric two-form background and the gauge field acquire
physical degrees of freedom unlike the D-string case. The three-form
field strength corresponding to the background field can be seen to
contribute towards the curvature on the world-volume.
In addition, we consider the appropriate super-partners in the present
context and study the $D_p$-brane world-volume geometry. We analyze the
the gauge theory on the $D_p$-brane world-volume and arrive at a
non-local description. It can be seen that the non-local fields are
essentially due to the presence of a generic antisymmetric background
field.{\footnote{
In particular, for a constant background field the non-locality disappears
and leads to a rotated $D_p$-brane.}}
We perform the analysis for the $D_p$-branes{\footnote{
`$p$' takes even value for type IIA and odd
in case of type IIB.}} in type IIA or IIB super-string theory 
where they play a natural role due to
their RR charges. For definiteness, we consider an arbitrary
$D_p$-brane ($p = 1,2,3,\dots $) in the
bosonic sectors (NS-NS and RR)
of type II closed string theories. 
It can be seen that the 
induced metric gives rise to curvature on the world-volume.
In case of type II super-string, the Kalb-Ramond (KR) potential 
in the NS-NS sector can also be seen to
induce additional curvature on the world-volume of a $D_p$-brane
and makes it a curved one.

\sp
We plan to present an explicit computation for an 
arbitrary bosonic $D_p$-brane in (oriented) open string theory.
The duality of the world-sheet can be used to interpret the open string
result in the type II closed string theories. 
We take into account the derivative expansion of the
collective coordinates next to leading order in $\a'$
and the effective dynamics can be seen to
be that of the Dirac-Born-Infeld (DBI). In fact,
the field contents in the NS-NS
sector of the type IIA or type IIB 
closed string and that of the oriented open bosonic
string are identical. The remaining fields in the RR sector act as sources
for the $D_p$-brane and determine its charges.
Obtaining the effective dynamics for the bosonic $D_p$-brane, we
generalize the formalism to a super-symmetric case by
considering appropriate fermionic partners for open string coordinates
as well as for the collective coordinates. To perform the
computations, we consider type I open super-string 
(un-oriented, $N=1$ super-symmetry) \cite{tseytlin,love},
and present necessary steps for the fermionic part to make the
presentation concise. In other words,
we investigate the world-volume dynamics of an arbitrary $D_p$-brane,
with the generic type II string backgrounds, in open string channel.
The presence of background fields can be
seen to be a deformation \cite{hulldo} of the $D_p$-brane world-volume
and the non-commutative geometry \cite{connesds,douglas1} shows up naturally
at the boundary of open string \cite{miaoli}. We investigate the 
induce geometry due to the non-zero and zero-modes of open string at 
its boundary and conclude with a note on the non-local description of 
gauge theory on the $D_p$-brane world-volume.

\section{Open string fluctuations on a p-brane}

It is known that the effective dynamics of a $D_p$-brane 
is due to the fluctuations governed
by the end of open super-string (whose world-sheet is a disk) with
appropriate boundary conditions. For instance, Dirichlet boundary
conditions in the transverse directions define the position of the 
$D_p$-brane and the remaining ($p+1$)-directions satisfy the Neumann 
conditions along the world-volume.

\sp

To begin with, consider an arbitrary $D_p$-brane
($p=1,2,\dots  9$), characterized by its space-time coordinates 
$f^{\mu}(t,\s_i )$, for ($i = 1,\dots  , p$) with Lorentzian signature 
($-,+,+,\dots +$), in presence of closed string 
background fields. The bosonic $D_p$-brane sub-manifold (world-volume) 
may be seen as an embedding in space-time
($\mu=0,1,\dots ,p, p+1,\dots $).
At the disk boundary{\footnote{For convenience,
we compute for the bosonic
string. Appropriate fermionic partners shall be introduced 
at a later point in section 8.}}
$\pr\S$, the Lorentz covariant condition becomes
\cite{dailp}
\be
X^{\mu}({\t})\ = \ f^{\mu}( t(\t ), {\s_i}({\t}) ) \ ,\label{dbdef}
\ee
where $X^{\mu}(z,{\bar z})$ denotes open string coordinates 
parameterized by the polar angle $\t$ with $(0\leq\t\leq2\pi )$ on $\pr\S$.

\sp
\hs
The interaction of the $D_p$-brane
with the massless excitations of closed string;
namely metric, $G_{\mu\nu}$, KR
antisymmetric two-form potential,
$B_{\mu\nu}$ and the dilaton, $\Phi$, can
be described in open string theory by a non-trivial generalization of 
our earlier formulations \cite{karkazama,kar} for a D-string. 
It is known that the open string possesses a $U(1)$ gauge field $A_{\mu}(X)$ 
in space-time and
can also be seen as a requirement for the consistency of closed
string KR potential \cite{witten}. The background
gauge field interacts with open string at its boundary while the
closed string background fields interact in bulk. Then the non-linear sigma 
model describing the dynamics of the open string can be written \cite{nappi} 
for a constant dilaton, in a conformal gauge, as
\bea
&& S[X,A,B ]\ = \ {1\over{4\pi\a'}} \ \int_{\S} d^2z \ \Big (\ G_{\mu\nu}(X)
{\pr_{\bar a}} X^{\mu}
{\pr^{\bar a}} X^{\nu}\ -\  i \ {\eps}^{\bar a\bar b} \ B_{\mu\nu}(X)
{\pr_{\bar a}} X^{\mu} {\pr_{\bar b}} X^{\nu} \ \Big ) \nonumber \\
&&{}\qquad\qquad\qquad\qquad\qquad\qquad\qquad\qquad
+ i \ {\int}_{\pr {\S}} d\t \ G_{\mu\nu}(X) A^{\mu} (X)
\ {\prt} X^{\nu} \ .\label{action}
\eea
The effective dynamics for the $D_p$-brane can be taken 
into account by generalizing the
path-integral \cite{tseytlin2,kazama,karkazama} in presence of curved 
background fields \cite{kar}. In fact, we make use of the 
Polyakov path-integral formulation \cite{polyakov}
to describe the generalized dynamics of the $D_p$-brane.
At low energy, the effective action for the $D_p$-brane can be obtained from
the renormalized non-linear sigma model partition function
which is the disk amplitude
modulo the massless closed string vertex. In this formalism, the
effective dynamics can be generalized from refs.\cite{karkazama,kar} and 
becomes
\be
{\cs }_{\rm eff}\Big ( f, A, B \Big ) =
{1\over{g_s}} \int \cd X^{\mu}(z,{\bar z}) \cd t(\t )
\ {\cd \s_i}({\t})\ {\d } \Big ( X^{\mu}(\t) - f^{\mu} 
(t(\t) , {\s_i}(\t)) \Big )
\cdot \exp \Big (-S[X,A,B ]\Big )\  , \label{path}
\ee
where $g_s= e^{\phi}$ is the closed string coupling constant. The  
$\d$-function in eq.(\ref{path}) takes care of the Dirichlet boundary
conditions and hence responsible for the $D_p$-brane description. The 
path-integral (\ref{path}) needs to be evaluated in bulk as well as at
the boundary, to obtain non-linear dynamics of a $D_p$-brane.

\section{Curvatures on the $D_p$-brane world-volume} 

The background fields can be seen to induce curvatures on the
$(p+1)$-dimensional world-volume of a $D_p$-brane.
The induced fields can be expressed in terms of the
$D_p$-brane collective coordinates, $f^{\mu}(t,\s_i )$, and become 
\bea
h_{ab} &=&  G_{\mu\nu}(X) \ {\pr }_a f^{\mu}{\pr }_b f^{\nu} \ ,
\nonumber \\
B_{ab} &=&
B_{\mu\nu}(X)\ {\pr }_a f^{\mu}{\pr }_b f^{\nu} \nonumber \\ 
{\rm and}\quad\quad
C_{abc\dots } &=& C_{\mu\nu\r\dots}(X)\ {\pr }_a f^{\mu}{\pr }_b f^{\nu}
{\pr }_c f^{\r} \dots \ ,
\eea
where $h_{ab}$, $B_{ab}$  and $C_{abc\dots }$ for 
($a,b,c, \dots =0,1,2\dots ,p$) 
are respectively the
metric, KR two-form and the RR ($p+1$)-form induced on the $D_p$-brane. 
This in turn leads to the
extrinsic curvature, $K$, on the $D_p$-brane world-volume.
The induced RR form mixes with the KR form in a gauge invariant 
way and the source term describes a 
($p+2$)-dimensional sub-manifold with its 
boundary representing the world-volume of the $D_p$-brane. For a generalized
$D_p$-brane, the boundary value of the Wess-Zumino (WZ) action 
can be given by \cite{douglas}
\be
S_{WZ}(f,C)\ =\ Q_p\ \int \ dt\ d^{p}\s\ e^{(B+{\bar F})} \wedge C \ ,
\label{charge}
\ee
where $Q_p$ denotes the p-brane charge density and the RR potential,
$C$, takes the appropriate forms with the $U(1)$ invariant field 
strength.{\footnote
{The effective action obtained by path-integrating eq.(\ref{path}) along
with the WZ action (\ref{charge}) play a vital role in the super-symmetric
formulation \cite{cederwall}.}}
The extrinsic curvature of a $D_p$-brane in presence of
arbitrary type II backgrounds depends on
the induced metric and KR two-form fields and can be expressed as
\bea
K^{\mu}_{ab}  &=&  P^{\mu\nu}\ \pr_a\pr_b f_{\nu}\ , \nonumber \\
K^{\mu}_{abc}  &= & \pr_a\pr_b\pr_c f^{\mu} - \pr_a {\G^{\mu}}_{bc}
+2 {\G^d}_{ab} {\G^{\mu}}_{dc} - {1\over2}
\pr_a {H^{\mu}}_{bc}
+{H^d}_{ab} {H^{\mu}}_{dc}
\ ,\label{ext}
\eea
where $\G$ is the Christoffel connection and $H\ (= dB)$ 
denotes the field strength
corresponding to the KR two-form.
$P^{\mu\nu}$ denotes the normal projected space and can be expressed in 
terms of the tangential space
$h^{\mu\nu}$:
\bea
&& P^{\mu\nu}=G^{\mu\nu} + {B}^{\mu\nu} - h^{\mu\nu} \ , \nonumber \\
{\rm where}\;\  && h^{\mu\nu}=\left (h^{ab} + B^{ab}\right )
\pr_af^{\mu}\pr_bf^{\nu} \ .
\label{projection}
\eea
In order to perform the boundary path-integrals, the brane coordinates
$f^{\mu}(t(\t),\s_i(\t))$ can be re-written in its derivative expansion
around its center of mass coordinate $f^{\mu}(t,\s_i)$. 
The appropriate geodesic expansion \cite{alvarez}
for the ($p+1$)-dimensional world-volume becomes 
\bea
f^{\mu}\pmatrix {{t(\t),\s_i(\t) } \cr }&=& f^{\mu}\big (t, \s_i\big ) \
+ \ {\pr }_a f^{\mu}\Big (t(\t ),\s_i({\t })\Big ) {\z }^a ({\t }) 
\ + \ {1\over2} K^{\mu}_{ab}
{\z }^a({\t }) {\z }^b({\t }) \nonumber \\
&&{}\qquad\qquad\qquad\qquad\quad
+ \ {1\over{3!}} K^{\mu}_{abc} {\z }^a({\t }) 
{\z }^b(\t ) {\z }^c(\t ) \ + \ {\co}({\z }^4 ) \ , \label{geodesic}
\eea
where ${\z }^a({\t })$ are the normal coordinates defined on the
world-volume of a p-brane.{\footnote{The expansion for a p-brane  
takes the similar form as 
the case of D-string \cite{karkazama}. However the explicit form of the
curvature (\ref{ext}) is much involved with geometry due to the
additional three-form field strength $H_{abc}$.}}

\sp
\hs
To leading order in the expansion (\ref{geodesic}),
the boundary of the disk is mapped on to a
point $(t,\s_i )$ on the ($p+1$)-dimensional world-volume. Then, an orthonormal
frame \cite{kazama,karkazama,kar}
can be set up at that point to simplify the computation. As a consequence,
the sub-leading terms in eq.(\ref{geodesic}), become the quantum fluctuations
in this frame. In addition, the basis vectors, $\e^{\mu}_a, \ (a=0,1,\dots 
p)$ span the ($p+1$)-dimensional tangential space and satisfy the Neumann 
boundary conditions. The remaining unit vectors, 
$\e^{\mu}_\alpha, \, (\a = p+1,\dots, 25)$
lie in a transverse space with Dirichlet boundary conditions. They can be
expressed for $A = (a, \a )$ as
\bea
\e^{\mu}_A &=& \cn_{(A)} \ e^{\mu}_A \ , \nonumber \\
\e^{\mu}_a &=& \cn_{(a)} \pr_a f^{\mu} \ , \nonumber \\
G^{\mu\nu} &=& {\e^{\mu}}_A {\e^{\nu}}_B \ \n^{AB} \ , \nonumber\\ 
B^{\mu\nu} &=& {\e^{\mu}}_A {\e^{\nu}}_B \ce^{AB} \ , \nonumber \\
h^{\mu\nu}  &=&  {\e^{\mu}}_a {\e^{\nu a}} 
\ + \ B^{ab} \ \e^{\mu}_a \e^{\nu}_b \ 
\nonumber\\
P^{\mu\nu} &=& {\e^{\mu}}_{\a} {\e^{\nu\a}} \ + \ B^{\a\b} \ \e^{\mu}_{\a}
\e^{\nu}_{\b} \ ,
\eea
where $\n_{AB}$ and $\ce_{AB}$ are the flat
backgrounds representing the Minkowskian metric and the KR two-form
respectively.
$\cn_{(a)}$ are the (induced) metric dependent normalizations in the
tangential space  and satisfy
\be
\sum_{(a\neq b)=0}^{p} \cn_{(a)}^{-2}\cn_{(b)}^{-2} \ =\ -\ h \ ,
\ee
where $h = \det h_{ab}$ and the
normalizations in the transverse space are denoted as
$\cn_{({\a})} = ( 1,1,\dots  1 )$.

\section{Boundary conditions for non-zero modes}

In order to simplify the computation, 
we separate out the zero modes, $x^{\mu}$,
from the string coordinates, $X^{\mu}(z, \bar z ) \ = \ x^{\mu} \ 
+ \ {\x}^{\mu} (z, \bar z )$. Then the non-zero modes, $\x^{\mu}$, can be
expressed in terms of the orthonormal coordinates 
$\r^A(z,\bar z)$:
\be
\x^{\mu}(z, \bar z )\ = \ \sum_A \e^{\mu}_A \ \r^A (z, \bar z ) \ ,
\ee
where $\r_a(z,\bar z ) $ and $\r_{\a}(z, \bar z ) $ correspond to the 
components in tangential
and transverse spaces respectively.

\sp
\hs
The background fields, namely metric $G_{\mu\nu}(X)$, 
KR two-form $B_{\mu\nu}(X)$ and
the $U(1)$ gauge field $A_{\mu}(X)$ can also be expanded
around their zero-modes:
\bea
G_{\mu\nu}(X)&=&\n_{\mu\nu} \ + \ \pr_{\l}G_{\mu\nu}\ \x^{\l}\ + \ \dots \ ,
\nonumber\\
B_{\mu\nu}(X)&=&\eps_{\mu\nu} \ + \ \pr_{\l}B_{\mu\nu}\ \x^{\l}\ + \ \dots 
\nonumber \\
{\rm and} \qquad\qquad A_{\mu}(X)&=&\ a_{\mu}\ +\ {1\over2}F_{\mu\nu}\ 
\x^{\nu}\ + \ \dots \ ,
\eea
where $\n_{\mu\nu}$ is the flat metric. $\eps_{\mu\nu}$ and
$a_{\mu}$ are the constant modes for the antisymmetric two-form and
gauge field respectively. For a constant
field strength $F_{ab}$ on the ($p+1)$-dimensional world-volume, the non-linear
sigma model action (\ref{action}) simplifies drastically in the orthonormal
frame and becomes \cite{kar}
\bea
&& S[\r,B,A] \ = \ {1\over{4\pi\a'}} \Big [\ - \int_{\S} d^2z \  
\r_A \pr^2 \r^A
\ +\ {\int}_{\pr\S} d\t \ \Big ( \ \r_A \pr_n \r^A \nonumber\\
&&{}\qquad\qquad\qquad\qquad\qquad\qquad\qquad 
+ \ i\ \cn_{(A)} \cn_{(B)}\  
\left (B_{AB} + {\bar F}_{AB} \ \right )\ \r^A
\pr_{\t} \r^B \ \Big )\ \Big ] 
\ ,\label{raction}
\eea
where $\pr_n$ denotes the normal derivative, 
$B_{AB} \equiv B_{\mu\nu} e^{\mu}_A e^{\nu}_B, 
{\bar F}_{AB} \equiv 2\pi\a' \ F_{AB}$,
$\pr_A \equiv e^{\nu}_A \nabla_{\nu}$ and  
$A_B \equiv e^{\mu}_B A_{\mu}$. 
Then the boundary conditions for the non-zero modes of the open string
can be derived from the
above eq.(\ref{raction}) and can be given as
\bea 
&&\pr_n \r_a (\t ) \ + \ \ i\ \cn_{(a)}\cn_{(b)} \
\left ( B_{ab} + \bar F_{ab} \right ) \ \pr_{\t}\r^b(\t ) \ = \ 0
\nonumber \\
{\rm and }\qquad \qquad && \qquad\r_{\a}(\t ) \ = \ 0 \ , \label{bcond}
\eea 
where $(B_{ab}+ {\bar F}_{ab})$ is the $U(1)$ invariant field strength on the
world-volume of an arbitrary $D_p$-brane.{\footnote{In section 7,
we analyze the boundary conditions to explain some of the interesting  
features of the world-volume dynamics.}}

\section{Integration over string modes}

\hs Now the effective dynamics of a
$D_p$-brane can be obtained by performing the
path-integrations in bulk as 
well as over the boundary fields. 
Let us consider the path-integral over the string modes $\r_A(z,\bar z)$ 
uniformly in bulk. In the orthonormal frame, the path-integral can be 
dealt separately
over the transverse and the longitudinal components. The path-integral
over the transverse components, $\r_{\a}$, is a Gaussian
and can be seen to be trivial using the Dirichlet boundary conditions 
(\ref{bcond}). Thus, the computation is essentially reduced to that 
over the longitudinal components $\r_a$. It can be expressed as
\bea
&& I_{\r}^L \ = \ \int \ \cd\r_a \  \exp\Big ( -{1\over{4\pi\a'}}\int_{\S} d^2z
\ \Big [\ \pr_{\bar a}\r^a \pr^{\bar a}\r_a \
-\ \cn_{(a)}\cn_{(b)}\Big ( B_{ab}
+ {\bar F}_{ab} \Big ) \nonumber \\
&& {}\qquad\qquad\qquad \cdot \d (|z| - 1) \ \r^a\pr_z\r^b \Big ] \ \Big )
\ \cdot \exp \Big ( \ i \int_{\S} d^2z\ \d ( |z| - 1 )
\ \nu_a(z)\ \r^a(z) \ \Big ) 
\ ,\label{string}
\eea 
where $\d(|z| - 1)\ \nu_a(z)= \nu_a(\t )$ denote the Lagrange multiplier fields
due to the boundary conditions in eq.(\ref{dbdef}).
The Gaussian integral (\ref{string}) is straight forward to perform and
one obtains 
\be
I_{\r}^L = \ \left [ {\rm Jacobian} \right ]\
\cdot \exp \Big ( \ {{\a'}\over2} \int d\t \ d\t' \ \cv_a(\t ) \
G_{ab}(\t , \t' ) \ \cv_b (\t' ) \ \n_{ab} \ \Big ) 
\ ,\label{bulk}
\ee
where $G_{ab}(\t, \t' )$ denotes the Neumann propagator on the boundary of
an unit disk. The matrix propagator satisfies
\bea
&& 
{}\qquad\qquad\qquad\qquad
\pr^2 G_{ab}(z,z')\ =\ 2 
\pi \ \n_{ab}\ \d^{(2)}(z,z') \qquad\qquad {\rm in\; bulk}
\nonumber\\
&&{\rm and}\;\;\;\; 
\pr_n G_{ab}(z,z') \ +\ i\ \cn_{(a)}\cn_{(b)}\ \left ( B_{ab}+{\bar F}_{ab}
\right )\ \prt G_{ab}(z,z')\ = \ 0 \qquad {\rm on}\;\; \pr\S \ .
\label{neumann}  
\eea
Finally, the expressions in eq.(\ref{neumann}) are analyzed and the
explicit form for the ($p+1$) dimensional square (orthogonal) 
matrix propagator can be expressed as

\bea
G_{ab}(z,z')\ = \ \n_{ab}\ \ln |z-z'|&+&{1\over2} \left ( {{1- \cn_{(a)}
\cn_{(b)}\left (B+{\bar F}\right )}\over{1+\cn_{(a)}\cn_{(b)}\left (B+{\bar F}
\right )}}\right )_{ab}\ \ln \left (1-{1\over{z{\bar z}'}}\right )
\nonumber \\
&& {}\qquad + \ {1\over2} \left ( {{1+ \cn_{(a)}
\cn_{(b)}\left (B+{\bar F}\right )}\over{1-\cn_{(a)}\cn_{(b)}\left (B+{\bar F}
\right )}}\right )_{ab}\ln \left (1-{1\over{z'{\bar z}}}\right ) \ .
\label{propagator}
\eea
The effect of the $U(1)$ gauge field can be seen as a Lorentz rotation
with respect to the one of vanishing gauge field. 
On the boundary $\pr\S$, the diagonal part of the propagator matrix
diverges as $\t\ra\t'$. We regularize the propagator by introducing a
cut off $\eps$ \cite{tseytlin}

\be
G_{aa}(\t ,\t')\ =\ -2\ \d_{aa}\ h\ \left [h -
\det\left (B_{ab}+{\bar F}_{ab}\right ) \right ]^{-1} \sum_{n=1}^{\infty}\
{{e^{-\eps n}}\over{n}}\cos \ n(\t -\t')\ .
\ee
In the limit $\t'\ \ra \ \t$, the propagator $G_{aa}(\t ,\t )$ can be
seen to contain a divergence and is given

\be
G_{aa}(\t ,\t)\ =\ 2\d_{aa} \ h \ \left [h - 
\det\left (B_{ab}+{\bar F}_{ab}\right )\right ]^{-1}\ \ln\ \eps \ .
\ee

\sp

\noindent
On the other hand, the Jacobian in eq.(\ref{bulk}) can be
written as  

\be
\left [ {\rm Jacobian} \right ] \ 
= \ {\left (- \det \left [ \n_{ab}\ \pr^2 \ + \ 
 \cn_{(a)}\cn_{(b)} \left ( B_{ab}
+ {\bar F}_{ab} \right )\d (|z|-1)\pr_z \right ]\ \right )}^{-{1\over2}} \ .
\ee
Using the Fourier mode expansion on a boundary circle, the Jacobian can  be
re-expressed as

\be
\left [ {\rm Jacobian}\right ]\  =\ \prod_{n=1}^{\infty} 
\left [ \ 1\ -\ \sum_{a,b=0}^p\ \cn_{(a)}^2 \cn_{(b)}^2\ 
\det\left (B_{ab}+{\bar F}_{ab}
\right )
\ \right ]^{-1} \ .
\ee

\noindent
The zeta-function regularization can be performed and one obtains
\bea
-\sum_{n=1}^{\infty} 1 &=&
- \lim_{q\ra 0} \sum_{n=1}^{\infty} n^{-q} \nonumber\\
&=& \z (0) \ .
\eea
Finally, the Jacobian for the path-integral in bulk reduces to a
simple form:
\be
\left [ {\rm Jacobian}\right ]\ = \ {1\over{\sqrt{h}}}\
\left ( \ h\ + B + \bar F
\ \right )^{1\over2} \ , \label{jacobian}
\ee
where $h+B+\bar F\ = \ \det (h_{ab}+B_{ab}+{\bar F}_{ab})$.
Since the computation of disk amplitude is obtained modulo for the massless
closed string modes, the Jacobian obtained (\ref{jacobian}) is
the only contribution from the bulk as a whole and corresponds to
a non-perturbative result (exact in $\a'$).

\section{Integration over boundary fields}

\hs
Now, we are in a position to perform the path-integral over the boundary
fields $\nu_a({\t})$ and $\z^a(\t )$ by assembling the relevant terms.
The path-integral over the Lagrange multiplier field, $\nu_a(\t )$, 
can be expressed as a Gaussian and thus straight-forward to
perform. The complexity arises in the computation are due to the curvatures 
and the source term can be explicitly expressed as
\bea
&& 
J_a(\t )\ =\ \cn_{(a)}^{-1}\ \n_{ab}\Big (\ \z^b(\t )\ -\ {1\over{3!}}\Big [\
K^{\mu}_{lm}K_{\mu np}\ h^{bp} \ + \ \G^b_{lp}\G^p_{mn}\ +\
 {1\over4} H^b_{lp}
H^p_{mn}\nonumber\\
&&{}\qquad\qquad\qquad\qquad 
+\ \pr_l\pr_m f^{\mu}\ \pr_qf_{\mu}\ \pr_nh^{bq}
+\ 2 \pr_l\pr_m f^{\mu}\pr_qf_{\mu}\G^q_{np}h^{bp}\nonumber \\
&&{}\qquad\qquad\qquad\qquad
+\ \pr_l\pr_m f^{\mu}\ \pr_q f_{\mu}\ H^q_{np}\ h^{bp}\
-\ \pr_rf^{\mu}\pr_qf_{\mu}\ \G^q_{np}\G^r_{lm}\ h^{bp}\nonumber\\
&&{}\qquad\qquad\qquad\qquad\quad
-{1\over4} \pr_rf^{\mu}\pr_qf_{\mu}\ H^q_{np}H^r_{lm}\ h^{bp}\ \Big ]
\ \z^l(\t)\z^m(\t )\z^n(\t )\ \Big ) \ + \ \co (\z^4 ) \ .
\label{source}
\eea
The result of the $\nu_a(\t )$-integration can be given
\be
I_{\nu} \ \equiv \ \exp\left ( {1\over{2\a'}}\int d\t\ d\t' J_a(\t )\
G_{ab}^{-1}(\t , \t' )\ J_{b}(\t' )\ \n_{ab} \right ) \ .
\ee

\sp
\hs
As can be analyzed from the source term (\ref{source}), to $\co(\a' )$,
the $\z_a(\t )$-integral contains a quartic interaction term apart from
the quadratic part and needs a perturbative treatment. 
In order to simplify the calculation, we generalize the
re-scaling for the D-string defined in previous refs.\cite{karkazama,kar} 
to an arbitrary p-brane. It becomes
\be
\z^a(\t )\ = \ {\sqrt{\a'}}\ \cn_{(a)} \ {\bar\z}^a(\t ) \ .
\ee
The corresponding change in the functional measure is calculated
by using the Fourier mode expansion and the zeta-function regularization.
Finally it can be expressed as
\be
\cd\z^a(\t ) \ =\ {\a'}^{-{{(p+1)}\over2}}\ 
\sum_{(a\neq b)=0}^{p} \cn_{(a)}^{-1} \ \cn_{(b)}^{-1}\ \cd {\bar \z}^a(\t )\ .
\ee

\hs
The quartic interaction term in the $\z^a(\t )$-integral can be 
simplified drastically by using the propagator
\be
{\bar\z}^a(\t ){\bar\z}^b(\t' ) \ = \ \n^{ab} \ G(\t, \t') \ .
\ee
After some calculations, the path-integrated boundary part
can be given by
\be
I_{\pr\S} \equiv {(\a' )}^{-{{(p+1)}\over2}}  {\sqrt{-h}}\
\left ( 1  - 
\a' \ h \left [ h-  
\det\left (B_{ab} + {\bar F}_{ab} \right )\right ]^{-1} 
\cn_{(a)}^2 \n_{aa} \
K^{\l}_{aa} \ K_{\l ab} \ h^{ab} \ln \eps + \co(\a'^2) \right )\ .\label{bound}
\ee

\sp
\hs
Now the effective dynamics of a $D_p$-brane can be obtained by considering
the path-integrated results in bulk (\ref{jacobian}) and boundary 
(\ref{bound}) along with the zero-modes contribution as a volume integral.
Considering all the factors properly, the disk amplitude
becomes
\bea
&&\cs_{\rm eff}(f,A,B) \ = \ 
T_p\ \int \ dt\ d^{p}\s \ {\sqrt{- \det\Big ( \ h + B +{\bar F}
\ \Big )}}
\nonumber\\
&&{}\qquad \cdot \left ( 1 - \a' \ h\
\left [ h - \det\left ( B_{ab} + {\bar F}_{ab} \right )
\right ]^{-1}
\cn_{(a)}^2
\ \n_{aa} \ K^{\l}_{aa} \ K_{\l ab}
h^{ab} \ \ln \eps \ +\ \co(\a'^2) \right ) \ , \label{evaluation}
\eea
where $T_p = 1/[ g_s(\a')^{{(p+1)}\over2}]$ 
denotes the $D_p$-brane tension. The above 
sub-leading term is due to
the extrinsic curvature, $K$, and is associated with a divergence 
($\ln \eps$). The 
singularity can be isolated by re-normalization of the string 
tension and does not affect the formulation. 
The corrections can be absorbed by a mass re-normalization which is also
the $D_p$-brane world-volume re-normalization. To
obtain a renormalized amplitude, we re-define the collective coordinates,
$f^{\mu}= f^{\mu}_R + \sum_a \pr_a f^{\mu}_R\ $, with a divergent piece:

\be
\d_af_R^{\mu} \ = \ - \ \a'\ \left ( {{h_R + \det\left (
B_{ab}^R + {\bar F}_{ab}\right ) }\over{h_R - \det\left (
B_{ab}^R + {\bar F}_{ab} \right )}}\right ) \ \n_{aa} \ \cn_{(a)}^2 
\ K^{\mu}_{aa} \ \ln \eps 
\ . \label{ren}
\ee
The index $R$ stands for the re-normalization of the $D_p$-brane.
Then the effective dynamics (\ref{evaluation}) for
a curved $D_p$-brane ($p=1,2,3\dots $), next to leading order, 
becomes precisely the Dirac-Born-Infeld
(DBI) action
\be
\cs_{\rm eff}(f_R,A,B_R) \ = \ 
T_{p} \ \int \ dt\ d^{p}\s \ 
{\sqrt{- \det \Big ( \ h_R \ + \ B_R \ + \ {\bar F} \ \Big )}}  \ +\
\co(\a'^2) \ . \label{dbi}
\ee
Thus the effective low energy dynamics of a curved $D_p$-brane, next to
leading order in its derivative expansion is described by the DBI action.
Since the computation is a low energy approximation, 
the renormalized DBI action (\ref{dbi}) receives corrections
from all the higher orders in $\a'$ which in turn is associated with
the derivative expansion of the $D_p$-brane coordinates, $f^{\mu}(t,\s_i )$,
in this formalism.

\section{String-modes and world-volume geometry }

In this section, we re-call the boundary conditions 
(\ref{bcond}), for the 
non-zero modes of open string, derived in section 4. The tangential
components of the string coordinates, $\r_a(\t )$, 
can be re-written and the boundary conditions for a $D_p$-brane
become
\bea
&&\pr_n {\r}_0 \ + \ i\ \cn_{(0)}\ \cn_{(i)} 
E_i\ \prt\r^i\ 
= \ 0 \ ,
\nonumber\\
&&\pr_n {\r}_i \ - \ i\ \cn_{(0)}\ \cn_{(i)} 
E_i\ 
\prt{\r}^0\ = \ 0 \nonumber \\
{\rm and}\qquad &&\pr_n {\r}_i \ + \ i\ \cn_{(i)}\ \cn_{(j)} 
(B_{ij} + {\bar F}_{ij})\ \prt\r^j\
= \ 0 \ ,
\label{bc}
\eea
where $E_i= (B_{0i}+{\bar F}_{0i})$ corresponds to the electric field
components $(E_1,E_2,\dots E_p)$ and
$(B_{ij}\ +\ {\bar F}_{ij})$ defines
the magnetic part of the background fields. For membranes
and higher dimensional branes, both 
electric and magnetic fields are non-vanishing unlike the string.
In an orthogonal moving frame,
the canonical momenta conjugate to ($\r^a, \r^{\a}$) can be 
expressed as
\bea
P^a(z,{\bar z})&=& \prt\r^a\ +\ i\cn_{(a)}\cn_{(b)}
\ (B^a_b +{\bar F}^a_b)\ \pr_n\r^b \nonumber\\
{\rm and}\qquad P^{\a}(z,{\bar z})&=& \prt\r^{\a} \ .\label{mom}
\eea
The equal time canonical commutators in bulk satisfy:
\bea
&& \Big [\ \r^a(z)\ ,\ \r^b(z')\ \Big ] = 0 \ ,\nonumber\\
&& \Big [\ P^a(z)\ ,\ P^b(z')\ \Big ] = 0 \nonumber \\
{\rm and}\;\ && \Big [\ \r^a(z)\ ,\ P^b(z')\ \Big ] = i \n^{ab}\ \d(z-z') \ .
\label{com}
\eea
The commutator (\ref{com}) can be simplified with the substitution from
the conjugate momenta
$P^b(z,{\bar z})$ in eq.(\ref{mom}) and turns out to be non-vanishing for
the non-zero modes in presence of the background fields. At the disk
boundary, the string fluctuations can be re-written in terms of the 
$D_p$-brane collective coordinates (non-zero modes)
in a static gauge and the equal time 
commutator becomes
\be
\Big [\ f^a(t(\t ),\s_i(\t))\ ,\ f^b(t(\t ),\s_i(\t ))\ \Big ]_{\pr\S}\ 
=\ \pm 2\pi\ i (M^{-1}[B+{\bar F}])^{ab}
\ , \label{ncom} 
\ee
\be
{\rm where}\qquad\quad M_{ab} = \cn_{(a)}^{-1}\cn_{(b)}^{-1}\ 
\n_{ab} - \cn_{(c)}^2[B+{\bar F}]^c_a\ [B+{\bar F}]_{cb} \ .
\ee
The right hand side in eq.(\ref{ncom}){\footnote{Interestingly, similar
notion for the coordinate field commutator can be also found in the
semi-classical frame-work \cite{verlinde}
of gravity coupled to SU(2) gauge theory 
\cite{karmaharana}. Since $(B+\bar F)_{ab}$ is gauge invariant, 
the gauge field can be absorbed in two-form field by Higgs mechanism
known in super-gravity. Then for large KR-field, the r.h.s. of the
commutator (\ref{ncom}) simplifies to $\pm 2\pi i B^{-1}$.}}
is due to the induced
fluxes on the world-volume of a $D_p$-brane leading to the non-commutative
geometry \cite{connesds,miaoli}. 
From the string theory perspective, the non-commutative geometry 
is a boundary phenomena and thus closed strings do not perceive this 
special geometry. However, for a $D_p$-brane the non-commutative feature
is in bulk which represents its world-volume.

\sp
Now generalizing the Lorentz rotation (R)-matrix from our earlier discussions 
\cite{kar}, the non-trivial components of the non-zero modes can be 
re-written as ${\tilde\r}^a = R^a_b\
\r^b$. Then, the boundary conditions for the
new $D_p$-brane, in terms of the transformed (tangential) coordinates, become
\bea
\pr_{e{\pm}}\ {\tilde\r}_i&=& 0 \nonumber\\
\pr_{m_p{\pm}}\ {\tilde\r}_{(p)i}&=& 0 \nonumber\\
{\rm and} \qquad\qquad {\r}_{\a} &=& 0 \ ,\label{mbc}
\eea
where $\pr_{e\pm} = [\pr_n \ \pm i\ \cn_{(0)}\cn_{(i)}\ E_i
\prt ]$ lie on the ($0i$)-planes and are defined with respect to
the boost from the original frame. 
The remaining components are magnetic in nature and are used to rotate 
the ($ij$)-planes:
$\pr_{m_p\pm}= [\pr_n \ \pm i\ \cn_{(i)}\cn_{(j)}
\ (B_{ij} + {\bar F}_{ij})\prt ]$. Here the
transformed boundary conditions for the new $D_p$-brane in the
tangential directions (${\tilde\r}_i, {\tilde\r}_{(p)i}$)
are the usual Neumann conditions. At a first sight, 
the conditions (\ref{mbc}) appear 
identical to a new $D_p$-brane in absence of 
electric and magnetic fluxes. Nevertheless, the fluxes are manifested
to rotate the original $D_p$-brane to a new one and 
the transformed tangential coordinates, ${\tilde\r}_a$, become non-local.
Thus the non-locality depends on electric and magnetic fluxes
and the explicit
R-matrix{\footnote{The modified Neumann matrix (\ref{propagator})
can be tuned to the usual disk propagator by the R-matrix at the expense 
of electric and magnetic fluxes.}} 
can be given by
\be
R_{ab}\ =\left ( {{1-\cn_{(a)}
\cn_{(b)}\left (B+{\bar F}\right )}\over{1+\cn_{(a)}\cn_{(b)}\left (B+{\bar F}
\right )}}\right )_{ab} \ .
\ee
It is interesting to note that for a constant KR field with a gauge choice,
the rotation matrix becomes fixed and the non-locality disappears. 
On the other hand, for the general case the equal time commutator 
for the non-zero modes becomes a delta function
along the remaining spatial coordinates. This provides an explicit
demonstration that the non-commutativity on
the world-volume of a $D_p$-brane can be manifested as the non-locality
on its world-volume. 

\sp 
Until now, we have discussed on the aspects of non-zero modes
in presence of KR background field. This is due to the fact that
the zero-modes have been separated out from the string coordinates and
the integral over the non-zero modes
$$\int_{\pr\S} \r^A(\t ) \ =\ 0 \ . $$
Now we are in a position to address the inherent features associated with
string zero-modes on the world-volume due to the KR-field. We argue that
the non-commutative or non-locality on the $D_p$-brane world-volume can be
seen due to the zero-modes in presence of KR-field. Such a two-form field
induces a $D_{(p-2)}$-brane charge density on the $D_p$-brane world-volume.
In fact, these lower dimensional branes can be seen to be
non-local along the new (two) Dirichlet directions on the original
$D_p$-brane world-volume due to the zero-modes in the theory.{\footnote{
For instance, see ref.(\cite{kar}) for the discussion on non-locality of 
D-instanton on D-string world-sheet.}} Thus there is evidence of no-local
description on the world-volume due to the KR-field along with the zero-modes.
Following the discussion for the non-zero modes in the previous paragraph,
one may view the non-locality as non-commutative on the $D_p$-brane
world-volume. In fact, the arbitrary value of the
commutator for the zero modes can be fixed by considering the string
(Fourier) mode expansion \cite{nappi} around the zero mode. In the context of
$D_p$-brane, a similar analysis can be performed to reveal the non-commutative
feature of zero-modes on the world-volume due to KR-field. Taking
the (Fourier) zero-modes into account, the
commutator (\ref{ncom}) for the collective coordinates  
can be re-defind in terms of the background fields and the non-commutative
geometry can be seen to sustain on the world-volume.

\section{Generalization to super $D_p$-branes}

The $D_p$-brane dynamics obtained 
(\ref{dbi}) in presence of an arbitrary string backgrounds
may be generalized to include the respective fermionic partners for the
open string coordinates $X^{\mu}(z,\bar z )${\footnote{Here the
space-time index is understood with $(\mu=0,1,2 \dots 9)$.}}
as well as for the collective coordinates $f^{\mu}(t,\s_i )$. 
In general, the super-symmetric $D_p$-brane analysis becomes highly
non-trivial in presence of super-string backgrounds, mostly due to the
$D_p$-brane collective coordinates, $f^{\mu}(t,\s_i)$, and the
$U(1)$ gauge field, $A^a$,
living on its world-volume. However, for a constant $U(1)$
gauge field strength $F_{ab}$, the computations simplify to some extent
in an orthonormal frame and the effective dynamics
for a super $D_p$-brane may be addressed with super-string background
fields. 

\sp
As a first step, towards a 
generalization of the $D_p$-brane effective dynamics to 
the case of the super $D_p$-brane, one needs to consider the type I 
super-string.{\footnote{In the 
bosonic case, we have considered oriented ($B\neq 0$)
open string to obtain the effective dynamics
of a $D_p$-brane. It allows one to analyze the type II
($N=2, D=10$) closed super-string channel and leads to non BPS $D_p$-brane.
However, a consistent open super-string
generalization reduces the super-symmetry to $N=1$ and makes $B=0$ (type I).
In this case, one finds BPS $D_p$-brane for ($p = 1,5,9 $).}}
In fact, type I open super-string is un-oriented and thus can
be considered as the interacting open and closed strings ($B=0$).
In this section, we briefly present the computations for the
fermionic partners by drawing analogy
from that of the bosonic case (with vanishing KR potential)
to avoid any repetition. This is indeed the case for $D_p$-branes
with $p=1,5,9 $ in type I theory.

\sp
Now re-consider the non-linear sigma model (\ref{action}) with
the Majorana fermions $\psi^{\mu}(z)$ as the
super-partners of the
(un-oriented) open string coordinates $X^{\mu}(z,\bar z )$. Also,
consider $\chi^{\mu}(t,\s_i )$ as the fermionic 
coordinates for the $D_p$-brane. For simplicity, the zero-modes are
separated out from the fermionic coordinates similar to the one
under the bosonic discussions. 
In an orthonormal frame, with a conformal gauge, the 
non-zero fermionic modes in the non-linear sigma model 
action{\footnote{We use the 
notation in ref.\cite{gsw}.}} in the Neveu-Schwarz-Ramond formalism
can be expressed as
\be
S(\psi , A)\ = \ {i\over{4\pi\a'}} 
\Big (\ - \ \int_{\S} d^2z \ 
\psi_A\pr\psi^A\  
+\ \cn_{(A)} \cn_{(B)}\
{\bar F}_{AB} \ {\int}_{\pr\S} d\t \
\ \psi^A\ \psi^B 
\Big )
\ ,\label{saction}
\ee
where $\psi^A=\e^A_{\mu}\psi^{\mu}$. Then the expression (\ref{saction}) along
with that in eq.(\ref{raction}) 
for vanishing KR potential ($B=0$),
takes care of the interacting type I open super-string with the massless
closed string fields. The complete non-linear sigma model 
action (eqs.(\ref{raction}) and 
(\ref{saction})) can be seen to be
invariant under the super-symmetric transformations 
\bea
&& \d\r^A\ =\ {\bar\eps}\psi^A \nonumber\\
{\rm and} \;\ && \d\psi^A\ =\ - i \eps \gamma^{\bar a}\pr_{\bar a}\r^A \ ,
\eea
where $\eps$ denotes an infinitesimal (constant) Majorana spinor and
$\gamma^{\bar a}$ denote two-dimensional matrices on the string
world-sheet. In addition, the covariant condition (\ref{dbdef})
is also modified due to the fermions at the disk boundary 
and can be expressed as a
constraint in the path-integral. 

\sp
Now the path-integral (\ref{path}) can be
generalized appropriately by taking into account the proper fermionic measures.
The effective dynamics for the non-zero modes can be described 
manifestly in super-field notations \cite{friedan} and in a static gauge
takes the form:
\be
{\hat{\cs }}( f, \chi )_{SUSY}  =
{1\over{g_s}} \int \ {\hat\cd} {\hat X}^A(z,\phi ) \ {\hat\cd} {\hat\s_a}(\t )
\ {\d } ({\hat X}^A - {\hat f}^A)_{\pr\S} \ 
\exp \Big (-{\hat S}[X,\psi ]_{SUSY}\Big )\  , \label{spath}
\ee
where $\hat\cd = \pr_{\phi} - \phi \pr_z$ denotes the super derivative
(${\hat\cd}^2=\pr_z$) and ($\phi , {\bar\phi}$) correspond to the 
super-symmetric
partners of ($z, \bar z $). The collective coordinates for the
super $D_p$-brane is represented by
${\hat f}^{\mu}(\s_a,\b_a )$, where $\b_a$ denotes the super-partners of 
$\s_a$. The super-space string coordinates become
\be
{\hat X}^A(z,\phi )= X^A(z)+\phi\ \psi^A(z)
\ee
and the corresponding measure in the path-integral (\ref{spath})
takes the form:
$$ \hat\cd {\hat\s}_a=
\cd \s_a \cd \b_a \ .
$$
The super-space action in eq.(\ref{spath}) can be expressed as
\be
{\hat S}[\r,\psi ]_{SUSY} = S[X,A,B]_{(B=0)} + S[\psi , A]
\ee
and corresponds to an
interacting type I super-string dynamics at low energy. It is straight
forward to note that the dynamics of a $D_p$-brane is due to the 
underlying type I 
super-string coupled to the curved backgrounds of closed string.
The additional boundary condition arises from the fermions in the
NS and Ramond sectors.
For the non-zero modes, they can be given by
\bea
&&\Big ( \psi_{+}^a \mp i\psi_{-}^a \Big )\ -\ \cn_{(a)}\cn_{(b)}\ 
{\bar F}^a_b \Big (\psi_{+}^b \pm i\psi_{-}^b\Big )\ =\ 0 \ 
\nonumber \\
{\rm and}\; \ && \Big ( \psi_+^{\a}\ \pm \ \psi_-^{\a}
\Big ) \ =\ 0 \ , \label{fbc}
\eea
where $\psi_{\pm}^A$ are the left and right moving
components of $\psi^A$ on the
string world-sheet. The first 
expression in eq.(\ref{fbc}) is the modified Neumann condition in 
presence background fields while the remaining one there
represents the Dirichlet boundary condition.

\sp
The induced super-symmetric invariant metric, ${\hat h}_{ab}$, and the RR
$(p+1)$-form, ${\hat C}_{abc\dots }$,  
are modified appropriately
due to the fermionic coordinates, $\chi^A(\s_a)$, in a static
gauge, on the world-volume of the
$D_p$-brane. The super-symmetric (invariant) fields can be expressed in
the orthogonal frame and they take the form
\bea
&& {\hat h}_{ab} = \n_{AB} D_a{\hat f}^A D_b{\hat f}^B \nonumber\\
{\rm and} &&  
{\hat C}_{abc\dots } = {\hat C}_{ABC\dots }D_a{\hat f}^A D_b{\hat f}^B
D_c{\hat f}^C\dots \ .\label{susyback}
\eea
Where the super collective coordinates and the super derivatives can
be explicitly given by
\bea
 &&{\hat f}^A(\s_a,\b_a ) = f^A(\s_a) + \b_a\chi^A(\s_a) \nonumber\\
{\rm and} && D_a= \pr_{\b_a} - \b_a \pr_a \ .\label{sderivative}
\eea

\sp
Now the path-integration over the fermionic modes ($\psi^a, \psi^{\a}$)
can be performed by extending the analysis for the bosonic case
discussed in previous chapters 5 and 6. The calculation essentially
reduces to that of Jacobian involving the determinants. The 
path-integral in bulk becomes Gaussian and the Jacobian due to 
the fermions in the NS-NS sector is found to be unity.
Since the field contents in the NS-NS sector turns out 
to be similar to that of the bosonic string, the Jacobian
for the bosons in this sector
becomes the one obtained (\ref{jacobian}) in the bosonic case
with vanishing KR potential.
On the other hand, in the RR sector the Jacobian due to the fermions exactly
cancels that of the bosons. There is no substantial contribution to
the Jacobian in the RR sector. Thus, the presence of fermions 
give a trivial contribution to the effective dynamics and do not affect 
the result obtained in bulk for the bosonic case. In fact, the 
super-string effective action takes the same form as the bosonic string
theory though the field contents are modified with respect to the
fermionic partners. 

\sp
On the other hand, the path-integral over the boundary fields 
involves the fermionic partners and has to be evaluated perturbatively. 
The non-triviality involves is mainly due to the
super $D_p$-brane collective coordinates ${\hat f}^{\mu}(\s_a(\t ),\b_a(\t ))$.
To calculate the boundary contribution, we consider the geodesic
expansion{\footnote{It can be expressed from that of the bosonic one
(\ref{geodesic}) by replacing the partial derivatives, $\pr_a$, with the
super-derivatives, $D_a$, as in eq.(\ref{sderivative}) 
and taking a note on the
super-coordinates, ${\hat f}^{\mu}(\s_a,\b_a)$.}}
about a point ($\s_a,\b_a$) in super-space \cite{friedan}. As a
consequence, the computations become non-trivial 
due to the modified induced fields (\ref{susyback}) on the world-volume.
The expression for the generalized extrinsic curvature, ${\hat K}$, can
be obtained readily from the modified fields. After a considerable
computations, the boundary path-integrals can be evaluated.{\footnote{
We skip the details of the steps involving re-normalization of the
super-brane coordinates, which can also be intuitively
obtained in hat notations from that of the bosonic case discussed.}}
The result of the
path-integrations in bulk as well as at the boundary can be encoded
in the generalized DBI action describing the super $D_p$-brane
($p=1,5,9$) dynamics
in open string theory. Finally it can be expressed as 
\be
{\hat\cs}(f,\chi )_{SUSY} \ = \
T_{p} \ \int \ d^{p+1}{\hat\s} \
{\sqrt{- \det \Big ( \ {\hat h} \ + \ {\bar F} \ \Big )}}  \ 
+\ Q_p\ \int\ d^{p+1}{\hat\s}\ e^{\bar F}\wedge {\hat C}\ . \label{sdbi}
\ee

\section{Discussion}

To summarize, the computations were performed in two parts, namely:
for the bosonic $D_p$-brane and then for the super $D_p$-brane. 
In the first part, 
we have demonstrated an explicit computations for an arbitrary
$D_p$-brane ($p=0,1,2,\dots $) in presence of the 
generic closed string backgrounds 
to obtain the non-linear dynamics in open bosonic string theory. It was
shown that the path-integration in bulk is essentially responsible for 
the non-perturbative dynamics. On the other hand, the integrations over the
boundary fields were performed next to leading order in $\a'$. The
perturbative corrections were expressed in terms of induced curvatures
on the world-volume of the $D_p$-brane and were found to be associated 
with a logarithmic divergence. A suitable re-normalization of the 
$D_p$-brane collective coordinates, $f^{\mu}(t,\s_i)$, were performed to
obtain the non-linear DBI action describing the
non-perturbative dynamics of a $D_p$-brane. In addition, the boundary
conditions for an arbitrary $D_p$-brane, in presence of close string 
backgrounds, were re-viewed with the zero-modes. In a static gauge,
the $D_p$-brane collective coordinates, $f^{\mu}(t,\s_i)$, were found to
be non-commutative due to the KR-field and the world-volume becomes 
non-local at the expense of the non-commutative geometry. The zero-modes
were analyzed in presence of the KR-field and a note on its
non-commutative feature is mentioned. Further analysis of the
world-volume geometry is believed to enhance the
understanding of gauge theory and would be addressed else where.

\sp
In the second part, our computations were generalized to include the
appropriate super partners in the type I super-string theory. The choice
of orthonormal moving frame facilitates the inclusion of fermions in the
locally inertial frame. 
The non-perturbative part of the super $D_p$-brane dynamics
was obtained by performing the Gaussian path-integration in
bulk. The boundary integrals were evaluated perturbatively and the
corrections in $\a'$ were 
found to be associated with the derivative expansion of the $D_p$-brane
super-coordinates, ${\hat f}^{\mu}(\s_a,\b_a)$. An qualitative analysis
for the boundary integrals were discussed by drawing analogy from
its bosonic counterpart.
Finally, the non-perturbative super $D_p$-brane dynamics were encoded in
the non-linear dynamics of the DBI action.

\sp
In this paper, we have computed the disk amplitude and thus the $U(1)$
gauge field on the world-volume is consistent. However, at the quantum
level (higher string loops), the consistency condition from the anomaly
cancellation of interacting open super-string would require the group to be
$SO(32)$. The issue of non-abelian world-volume dynamics becomes
technically difficult and some attempts can be found in 
ref.\cite{tseytlin3}.
Another important issue is the special world-volume
geometry due to the string-modes. From the string theory point of view,
this is a boundary phenomena. However for a $D_p$-brane, the non-commutative
feature is in the bulk of its world-volume. Since the world-volume theory
for a $D_p$-brane is a dimensionally reduced super Yang-Mills
to ($p+1$)-dimensions, the gauge theory becomes a non-local field theory.
This is a subtle issue and needs deep understanding of the subject.

\sp
\sp
\sp
\sp

{\large \bf Acknowledgments:} 
\sp

\hs
I wish to thank the members of the theory group here in the Institute for 
various discussions on the subject in general. The work is supported 
by the Swedish Natural Science Research Council.

\def\anp{Ann. of Phys.}
\def\prl{Phys. Rev. Lett.}
\def\prd#1{{Phys. Rev.} {\bf D#1}}
\def\plb#1{{Phys. Lett.} {\bf B#1}}
\def\npb#1{{Nucl. Phys.} {\bf B#1}}
\def\mpl#1{{Mod. Phys. Lett} {\bf A#1}}
\def\ijmpa#1{{Int. J. Mod. Phys.} {\bf A#1}}
\def\rmp#1{{Rev. Mod. Phys.} {\bf 68#1}}

\vfil\eject


\begin{thebibliography}{99}

\bibitem{notes}J. Polchinski, TASI Lectures on D-branes, {\tt hep-th/9611050};
C. Bachas, Lectures on D-branes, {\tt hep-th/9806199}.

\bibitem{polchinski}J. Polchinski, {\prl } {\bf 75} (1995) 4724
({\tt hep-th/9510017}).

\bibitem{callan}I. R. Klebanov and L. Thorlacius, \plb{371} (1996) 51
({\tt hep-th/9510200}); C. Bachas, \plb{374} (1996) 37 ({\tt hep-th/9511043});
C. G. Callan and I. R. Klebanov, \npb{465} (1996) 473; ({\tt hep-th/9511173});
C. Schmidhuber, \npb{467} (1996) 146 ({\tt hep-th/9601003});
M. R. Garousi and R. C. Myers, \npb{475} (1996) 193 ({\tt hep-th/9603194});
W. Fischler, S. Paban and M. Rozali, \plb{381} (1996) 62 
({\tt hep-th/9604014});
A. Hashimoto and I. R. Klebanov, \plb{381} (1996) 437 ({\tt hep-th/9604065});
C. Callan and J. Maldacena, \npb{513} (1998) 198 ({\tt hep-th/9708147});
M. R. Garousi and R. C. Myers, \npb{542} (1999) 73 ({\tt hep-th/9809100}).

\bibitem{dailp}J. Dai, R. G. Leigh and J. Polchinski, \mpl{4} (1989) 2073.

\bibitem{leigh}R. G. Leigh, \mpl{4} (1989) 2767.

\bibitem{mrdouglas}M. R. Douglas, J.Geom.Phys. {\bf 28} (1998) 255
({\tt hep-th/9604198}).

\bibitem{tseytlin3}A. A. Tseytlin, \npb{501} (1997) 41
({\tt hep-th/9701125}).

\bibitem{tseytlin4}A. A. Tseytlin, \npb{524} (1998) 41 ({\tt hep-th/9802133}).

\bibitem{kazama}S. Hirano and Y. Kazama, \npb{499} (1997) 495
({\tt hep-th/9612064}); Y. Kazama, \npb{\bf 504} (1997) 285
({\tt hep-th/9705111)}.

\bibitem{karkazama}S. Kar and Y. Kazama, \ijmpa{14} (1999) 1531
({\tt hep-th/9807239}).

\bibitem{kar}S. Kar, \npb{554} (1999) 163 ({\tt hep-th/9812230}).

\bibitem{cederwall}M. Cederwall, A. von Gussich, B. E. W. Nilsson, P. Sundell
and A. Westerberg, \npb{490} (1997) 179 ({\tt hep-th/9611159});
E. Bergshoeff and P. K. Townsend, \npb{490} (1997) 145 ({\tt hep-th/9611173});
J. Aganagic, C. Popescu and J. H. Schwarz, \npb{\bf495}
(1997) 99 ({\tt hep-th/9612080}).

\bibitem{tseytlin}E. S. Fradkin and A. A. Tseytlin, \plb{160} (1985)
69; A. A. Tseytlin, \npb{276} (1986) 391. 

\bibitem{love}C. G. Callan, C. Lovelace, C. R. Nappi and S. A. Yost,
\npb{308} (1988) 221.

\bibitem{hulldo}M. R. Douglas and C. Hull, J. High Energy Phys.
{\bf 9802} (1998) 003 ({\tt hep-th/9711165}).

\bibitem{connesds}A. Connes, M. R. Douglas and A. Schwarz, J.High Energy
Phys. {\bf 2} (1998) 003 ({\tt hep-th/9711162}).

\bibitem{douglas1}M. R. Douglas, {\tt hep-th/9901146} (1999).

\bibitem{miaoli}M. Li, {\tt hep-th/9802052}; 
M. Berkooz, \plb{430} (1998) 237 ({\tt hep-th/9802069});
Y.-K. E. Cheung and M. Krogh, \npb{528} (1998) 185 ({\tt hep-th/9803031});
F. Ardalan, H. Arfaei and M. M. Sheikh-Jabbari, J.High Energy Phys.
{\bf 9902} (1999) 016
({\tt hep-th/9810072}); C. Hofman and E. Verlinde, J.High Energy Phys.
{\bf 9812} (1998) 010
({\tt hep-th/9810116}); C. Chu and P. Ho, \npb{550} (1991) 151 
({\tt hep-th/9812219}); M. Kato and T. Kuroki, J.High Energy Phys.
{\bf 9903} (1999) 012 ({\tt hep-th/9902004}).

\bibitem{witten}E. Witten, \npb{460} (1996) 335 ({\tt hep-th/9510135}).

\bibitem{nappi}A. Abouelsaood, C. G. Callan, C. R. Nappi and
S. A. Yost, \npb{280} [FS18] (1987) 599;
C. G. Callan, C. Lovelace, C. R. Nappi and S. A. Yost,
\npb{288} (1987) 525.

\bibitem{tseytlin2}E. S. Fradkin and A. A. Tseytlin, \plb{163}
(1985) 123.

\bibitem{polyakov}A. M. Polyakov, \plb{103} (1981) 207.

\bibitem{douglas}M. R. Douglas, {\tt hep-th/9512077} (1999).

\bibitem{alvarez}L. Alvarez-Gaume, D. Z. Freedman and S. Mukhi, Ann.of Phys.
(1981) 85; E. Braaten, T. L. Curtright and C. K. Zachos, \npb{260} (1985) 630.

\bibitem{verlinde}H. Verlinde and E. Verlinde, \npb{371} 246 (1992),
({\tt hep-th/9110017}).

\bibitem{karmaharana}S. Kar and J. Maharana, \ijmpa{10} (1995) 2733, 
({\tt hep-th/9412026}).

\bibitem{gsw}M. B. Green, J. H. Schwarz and E. Witten, Superstring theory,
Vol. I, Cambridge Univ. Press (1987).

\bibitem{friedan}D. Friedan, E. Martinec and S. Shenkeer, \npb{271} (1986)
93.

\end{thebibliography}
\end{document}